\documentclass{jnmp01b}

\numberwithin{equation}{section}

\theoremstyle{plain}
 \newtheorem{proposition}{Proposition}[section]
 \newtheorem{theorem}[proposition]{Theorem}
\theoremstyle{definition}
 \newtheorem{definition}[proposition]{Definition}
 \newtheorem{remark}[proposition]{Remark}
 \newtheorem{example}[proposition]{Example}
 \newtheorem{property}[proposition]{Property}

\begin{document}

\renewcommand{\evenhead}{M.\ Cr\^a\c sm\u areanu}
\renewcommand{\oddhead}{First Integrals Generated by Pseudosymmetries in 
Nambu-Poisson Mechanics}


\thispagestyle{empty}

\begin{flushleft}
\footnotesize \sf
Journal of Nonlinear Mathematical Physics \qquad 2000, V.7, N~2,
\pageref{firstpage}--\pageref{lastpage}.
\hfill {\sc Letter}
\end{flushleft}

\vspace{-5mm}

\copyrightnote{2000}{M.\ Cr\^a\c sm\u areanu}

\Name{First Integrals Generated by Pseudosymmetries in Nambu-Poisson 
Mechanics}

\label{firstpage}

\Author{Mircea CR\^A\c SM\u AREANU}

\Adress{Faculty of Mathematics, University ``Al. I. Cuza'', Ia\c si, 6600,
Romania \\
E-mail: mcrasm@uaic.ro \\
and\\
Institute of Mathematics ``Octav Mayer'', Ia\c si Branch of
Romanian Academy, \\
Ia\c si, 6600, Romania}

\Date{Received November 2, 1999; Revised December 24, 1999; 
Accepted January 28, 2000}

\begin{abstract}
\noindent
Some types of first integrals for Hamiltonian Nambu-Poisson vector fields
are obtained by using the notions of pseudosymmetries. In this theory, the
homogeneous Hamiltonian vector fields play a special role and we point out this
fact. The differential system which describe the $SU(2)$-monopoles is given
as example. The paper ends with two appendices.
\end{abstract}


\section{Introduction}

Some physical systems can be described by using multibrackets instead of
Poisson brackets; for example: the Euler equations of the free rigid body,
the Maxwell-Bloch equations of laser-matter dynamics, the Toda lattice
equation and the heavy top. Beginning with Nambu \cite{n:m} and following
with Takhtajan \cite{t:n} these multibrackets, usually called {\it %
Nambu-Poisson structures}(NP structures, on short) led to some interesting
mathematical developments \cite{ch:d}, \cite{i:np}, \cite{v:l}, \cite{v:s}.

However, to the best of our knowledge, one of the main subjects of dynamics,
namely {\it the first integrals}, was not intensively studied except \cite
{ch:d}. The aim of the present paper is to make an attempt of studying this
subject. More precisely, two types of first integrals of Hamiltonian NP
vector fields are obtained by using the notions of pseudosymmetry and
adjoint pseudosymmetry. Also, the fact that the bracket of $n$-first
integrals is again a first integral is generalized.

The paper is structured as follows: the first section covers the main result
of the author's paper \cite{c:p}, the second section gives the basics of NP
theory and the third section presents applications to homogeneous
Hamiltonian vector fields on ${I\!\!R}^n$. The next section contains the
adjoint pseudosymmetries approach to find first integrals in NP dynamics.
Two appendices are inserted at the end of paper.

\section{First integrals generated by pseudosymmetries}

Since evolution in NP mechanics is given by a vector field, called {\it %
Hamiltonian vector field} like in symplectic mechanics, we will be
interested in methods to find first integrals for vector fields. In \cite
{c:p} we obtained such a method by using pseudosymmetries and this section
recall the results of the cited paper.

Let $M$ be a smooth, $m$-dimensional manifold, $C^\infty \left( M\right) $
the ring of real-valued smooth functions, ${\cal X}\left( M\right) $ the Lie
algebra of vector fields and $\Omega ^p\left( M\right) $ the $C^\infty
\left( M\right) $-module of $p$-differential forms, $1\leq p\leq m$. For $%
X\in {\cal X}\left( M\right) $ with local expression $X=X^i\left( x\right)
\frac \partial {\partial x^i}$ one considers the system of differential
equations which gives the flow of $X$: 
\begin{equation}
\dot{x}^i\left( t\right) =\frac{dx^i}{dt}\left( t\right) =X^i\left(
x^1\left( t\right) ,\ldots ,x^m\left( t\right) \right) .
\end{equation}
A solution of $\left( 2.1\right) $ is called an {\it integral curve }of $X$.

\begin{definition} 
A function ${\cal F}\in C^\infty \left( M\right) $ is
called a {\it conservation law }(or {\it first integral, }or {\it constant
of motion, }or {\it invariant function}){\it \ } of $X$ or of $\left(
2.1\right) $ if ${\cal F}$ is constant along the solutions of $\left(
2.1\right) $ that is 
\[
\frac{d\left( {\cal F}\circ c\right) }{dt}\left( t\right) =0 
\]
for every integral curve $c\left( t\right) $ of $X$.
\end{definition}

The following characterization is useful:

\begin{proposition} 
${\cal F}\in C^\infty \left( M\right) $ {\it is a
first integral of }$\left( 1.1\right) $ {\it if and only if } 
\[
{\cal L}_X{\cal F}=0 
\]
{\it where the right-hand side means} {\it the Lie derivative of }${\cal F}$%
{\it \ with respect to }$X$.
\end{proposition}

For our approach we need the following:

\begin{definition} 
(i) $Y\in {\cal X}\left( M\right) $ is called a {\it %
symmetry of }$X$ if 
\[
{\cal L}_XY=0.
\]
(ii) If $Y\in {\cal X}\left( M\right) $ is fixed then $Z\in {\cal X}\left(
M\right) $ is called a $Y${\it -pseudosymmetry of }$X$ if there exists $f\in
C^\infty \left( M\right) $ such that 
\[
{\cal L}_XZ=fY.
\]
(iii) $\omega \in \Omega ^p\left( M\right) $ is called an {\it invariant $p$%
-form of }$X$ if 
\[
{\cal L}_X\omega =0.
\]
If $p=1$ then $\omega $ is called an {\it adjoint symmetry} in \cite[p. 36]
{e:s}.
\end{definition}

The result which give the association between pseudosymmetries and first
integrals is:

\begin{proposition} 
{\it Let }$X\in {\cal X}\left( M\right) $ {\it be a
fixed vector field and }$\omega \in \Omega ^p\left( M\right) $ {\it be an
invariant}
$p-${\it form of }$X${\it . If } $Y\in {\cal X}\left( M\right) $ {\it is a
symmetry of }$X$ {\it and }$S_1,\ldots ,S_{p-1}\in {\cal X}\left( M\right) $ 
{\it are }$\left( p-1\right) $ $Y${\it -pseudosymmetries of }$X$ {\it then: }
\begin{equation}
{\cal F}=\omega \left( S_1,\ldots ,S_{p-1},Y\right) 
\end{equation}
{\it is a first integral for }$X$. {\it In particular, if }$Y,S_1,\ldots
,S_{p-1}$ {\it are symmetries of }$X$ {\it then }${\cal F}$ {\it given by }$%
\left( 2.2\right) $ {\it is a first integral.}
\end{proposition}

\begin{remark} 
(i) For $Y=X$ one obtain the main result of G. L. Jones 
\cite[p. 1056]{j:s}. \\ (ii) If $p=1$ one obtain Theorem 2.5.10 of ten
Eikelder \cite[p. 48]{e:s}. \\ (iii) Some applications to Lagrangian and
Hamiltonian systems can be found in \cite{c:p}. In these cases we have an
invariant $2$-form, the so-called {\it Cartan }$2${\it -form}. \\ (iv) In 
\cite{u:bf} it is solved a somewhat inverse problem: more precisely is given
a way to obtain a symmetry of a vector field which admits $n-1$ first
integrals. Let us note that \"{U}nal's method has tangency with the
Bayen-Flato generalization of Nambu mechanics.
\end{remark}

\begin{example}[Static SU(2)-monopoles] 
The Nahm's system in the theory of static SU(2)-monopoles is presented in 
\cite{i:np}: 
\begin{equation}
\frac{dx^1}{dt}=x^2x^3,\quad \frac{dx^2}{dt}=x^3x^1,\quad \frac{dx^3}{dt}%
=x^1x^2.
\end{equation}

The vector field $X=x^2x^3\frac \partial {\partial x^1}+x^3x^1\frac \partial
{\partial x^2}+x^1x^2\frac \partial {\partial x^3}$ is homogeneous of order
two, that is :
\begin{equation}
\left[ \Upsilon ,X\right] =X
\end{equation}
where :
\[
\Upsilon =x^1\frac \partial {\partial x^1}+x^2\frac \partial {\partial
x^2}+x^3\frac \partial {\partial x^3}. 
\]
But $\left( 2.4\right) $ means that $\Upsilon $ is $X$-pseudosymmetry for $%
\left( 2.3\right) $.
\end{example}

\section{Nambu-Poisson revisited}

In this section we review the necessary facts from NP structures. For more
details see \cite{ch:d}, \cite{i:np}, \cite{s:e1}, \cite{s:e2}, \cite{t:n}, 
\cite{v:l}, \cite{v:s} and the references therein.

\begin{definition} 
A {\it Nambu-Poisson bracket or structure} of order $n$%
, $2\leq n\leq m$ is an internal $n$-ary operation on $C^\infty (M)$,
denoted by $\{\;\}$, which satisfies the following axioms:\\(i) $\{\;\}$ is $%
I\!\!R$-multilinear and skew-symmetric\\(ii) the {\it Leibniz rule:} 
\[
\break \{f_1,\ldots ,f_{n-1},gh\}=\{f_1,\ldots ,f_{n-1},g\}h+g\{f_1,\ldots
,f_{n-1},h\} 
\]
(iii) the {\it fundamental identity:} 
\[
\{f_1,\ldots ,f_{n-1},\{g_1,\ldots ,g_n\}\}=\sum_{k=1}^n\{g_1,\ldots
,\{f_1,\ldots ,f_{n-1},g_k\},\ldots ,g_n\}. 
\]
Remember that if we use the same definition for $n=2$, we get a {\it Poisson
bracket}.
\end{definition}

By (ii), $\{\;\}$ acts on each factor as a vector field, whence it must be
of the form: 
\[
\{f_1,\ldots ,f_n\}=\Lambda (df_1,\ldots ,df_n) 
\]
where $\Lambda $ is a field of $n$-vectors on $M$. If such a field defines a
NP bracket, it is called a {\it NP tensor (field)}. $\Lambda $ defines a
bundle mapping 
\[
\sharp _\Lambda :\underbrace{T^{*}M\times \ldots \times T^{*}M}%
_{(n-1)\ \mathrm{times}}\longrightarrow TM 
\]
given by: 
\[
<\beta ,\sharp _\Lambda (\alpha _1,\ldots ,\alpha _{n-1})>=\Lambda (\alpha
_1,\ldots ,\alpha _{n-1},\beta ) 
\]
where all the arguments are covectors.

The next basic notion is that of the $\Lambda $-{\it Hamiltonian vector field%
} of $(n-1)$ functions defined by: 
\[
X_{F_1\ldots F_{n-1}}=\sharp _\Lambda (dF_1,\ldots ,dF_{n-1}). 
\]

Related to our subject of interest, namely first integrals, there are three
important properties:

\begin{property} 
A function $f\in C^\infty \left( M\right) $ is a first
integral of $X_{F_1\ldots F_{n-1}}$ if and only if 
\[
\{F_1,\ldots ,F_{n-1},f\}=0. 
\]
\end{property}

\begin{property} 
The NP bracket of $n$ first integrals is again a first
integral.
\end{property}

\begin{property}
The Hamiltonian vector fields are infinitesimal
automorphisms of the NP tensor: 
\[
L_{X_{F_1\ldots F_{n-1}}}\Lambda =0. 
\]
\end{property}

\section{The case $M\subseteq $ ${I\!\!R}^n$}

Let us suppose that $M\subseteq {I\!\!R}^n$(this imply $n=m$ !) and introduce
 the following differential forms:

(i) $\Omega =dx^1\wedge \ldots \wedge dx^n$, the volume $n$-form,

(ii) $\omega _{F_1\ldots F_{n-1}}=i_{X_{F_1\ldots F_{n-1}}}\Omega
=dF_1\wedge \ldots \wedge dF_{n-1}$, the $\left( n-1\right) $-form
associated to $X_{F_1\ldots F_{n-1}}$( cf. \cite{m:l}).

The lemma 1 from \cite{m:l} gives that $\Omega $ is an
invariant $n$-form of $X_{F_1\ldots F_{n-1}}$ or in other words $%
X_{F_1\ldots F_{n-1}}$ is a {\it solenoidal} vector field. Also, from $%
i_X^2=0$ it results that $\omega _{F_1\ldots F_{n-1}}$ is an invariant $%
\left( n-1\right) $-form of $X_{F_1\ldots F_{n-1}}$. Therefore in the NP
setting on $M\subseteq {I\!\!R}^n$ there are two invariant forms for a
Hamiltonian vector field. Then we have the following form of Proposition 2.4:

\begin{proposition} 
{\it If }$Y$ {\it is a symmetry of }$X_{F_1\ldots
F_{n-1}}$ {\it and }$S_1,\ldots ,S_{n-1}$ {\it are }$n-1$ $Y${\it %
-pseudo\-symme\-tries of }$X_{F_1\ldots F_{n-1}}$ {\it then:} 
\[
{\cal F}_1=\Omega \left( Y,S_1,\ldots ,S_{n-1}\right) 
\]
\[
{\cal F}_2=\omega _{F_1\ldots F_{n-1}}\left( Y,S_1,\ldots ,S_{n-2}\right) 
\]
{\it are first integrals of }$X_{F_1\ldots F_{n-1}}${\it . In particular:}

(i) $n=2,$ ${\cal F}_2=\omega _{F_1}\left( Y\right) =Y\left( F_1\right) $

(ii) $n=3,{\cal F}_2=dF_1\wedge dF_2\left( Y,S_1\right) =Y\left( F_1\right)
S_1\left( F_2\right) -Y\left( F_2\right) S_1\left( F_1\right) $.
\end{proposition}

In the following let us suppose that $X_{F_1\ldots F_{n-1}}$ is $p$%
-homogeneous, that is \linebreak[4]
$\left[ \Upsilon ,X_{F_1\ldots F_{n-1}}\right] =\left(
p-1\right) X_{F_1\ldots F_{n-1}}$ 
where $\Upsilon =x^1\frac \partial
{\partial x^1}+\ldots +x^n\frac \partial {\partial x^n}$ and $p$ is an
integer. In this case applying the previous result with $Y=X_{F_1\ldots
F_{n-1}}$ and $S_{n-1}=\Upsilon $ we get:

\begin{proposition} 
{\it Let $X_{F_1\ldots F_{n-1}}$ be a $p$-homogeneous
Hamiltonian vector field on $M\subseteq {I\!\!R}^n$. If $S_1,\ldots ,S_{n-2}$
are $n-2$ $X_{F_1\ldots F_{n-1}}$-pseudosymmetries of $X_{F_1\ldots F_{n-1}}$
then: 
\[
{\cal F}_1=\omega _{F_1\ldots F_{n-1}}\left( \Upsilon ,S_1,\ldots
,S_{n-2}\right) 
\]
is a first integral of }$X_{F_1\ldots F_{n-1}}${\it . In particular, if }$n=2
$ {\it then }${\cal F}_1=\omega \left( \Upsilon \right) $ {\it is a first
integral of }$X_{F_1}${\it .}
\end{proposition}

\begin{trivlist}\item[]
{\bf Example 2.6 revisited.}\ 
The system $\left( 2.3\right) $ can be written in NP form with(\cite{i:np}): 
\begin{equation}
\{f_1,f_2,f_3\}=\frac{\partial \left( f_1,f_2,f_3\right) }{\partial \left(
x^1,x^2,x^3\right) }
\end{equation}
and: 
\begin{equation}
F_1=\frac 12\left( \left( x^1\right) ^2-\left( x^2\right) ^2\right) ,\quad
F_2=\frac 12\left( \left( x^1\right) ^2-\left( x^3\right) ^2\right) .
\end{equation}
\end{trivlist}

So, because this system is NP and homogeneous, the previous proposition
works. For a list of others homogeneous Hamiltonian vector fields see
Appendix 1.

We return to the case $n=2$ of Proposition 4.1 for local expressions.
So, let $\{,\}$ be a Poisson bracket on ${I\!\!R}^2$ and $F\in C^\infty
\left( {I\!\!R}^2\right) $. Then: 
\[
X_F=\frac{\partial F}{\partial x^2}\frac \partial {\partial x^1}-\frac{%
\partial F}{\partial x^1}\frac \partial {\partial x^2} 
\]
\[
\omega _F=dF=\frac{\partial F}{\partial x^1}dx^1+\frac{\partial F}{\partial
x^2}dx^2 
\]
\[
{\cal F}_1=\omega _F\left( \Upsilon \right) =\frac{\partial F}{\partial x^1}%
x^1+\frac{\partial F}{\partial x^2}x^2={\cal L}_\Upsilon F 
\]
and the $p$-homogeneity of $X_F$ reads as follows: 
\[
{\cal L}_\Upsilon \frac{\partial F}{\partial x^i}=p\frac{\partial F}{%
\partial x^i},\quad 1\leq i\leq 2. 
\]

Therefore we get:

\begin{proposition} 
{\it Let }$\{,\}$ {\it be a Poisson bracket on } ${%
I\!\!R}^2$ {\it and }$F\in C^\infty \left( {I\!\!R}^2\right) $ {\it such
that the functions }$\frac{\partial F}{\partial x^i},1\leq i\leq 2$ {\it are 
}$p${\it -homogeneous. Then }${\cal F}_1={\cal L}_\Upsilon F$ {\it is a
first integral of the Hamiltonian vector field }$X_F${\it .}
\end{proposition}

At the beginning of this section we use the P. Morando's result that every NP
Hamiltonian vector field on ${I\!\!R}^n$ is solenoidal. In the following we
give a local converse of this topic.

So, if $X\in {\cal X}\left( {I\!\!R}^n\right) $ is solenoidal, a generalized
Euler theorem states that there are $F_1,\ldots ,F_{n-1}$ smooth functions
in a neighbourhood $U$ of every regular point $p$ of $X$(that is $X\left(
p\right) \neq 0$) such that: 
\begin{equation}
X=\nabla F_1\times \ldots \times \nabla F_{n-1}
\end{equation}
where $\nabla F$ means the gradient of $F\in C^\infty \left( U\right) $ and $%
\times $ is the vector product of ${I\!\!R}^n$. But $\left( 4.3\right) $ is
exactly the expression Nambu-Poisson of $X$.

\section{From adjoint pseudosymmetries to first integrals in\\ Nambu-Poisson
dynamics}

In order to obtain a result similar to Proposition 2.4 in NP mechanics we 
need:

\begin{definition} 
Let $X\in {\cal X}\left( M\right) $ and $\omega \in
\Omega ^1\left( M\right) $. Then a given $\alpha \in \Omega ^1\left(
M\right) $ is called an $\omega $-{\it adjoint} {\it pseudosymmetry} of $X$
if 
\[
{\cal L}_X\alpha =\rho \omega 
\]
for some $\rho \in C^\infty \left( M\right) $.
\end{definition}

A straightforward computation gives another main result of this paper:

\begin{proposition} 
{\it Let }$X_{F_1\ldots F_{n-1}}$ {\it be a
Hamiltonian vector field and }$\omega \in \Omega ^1\left( M\right) $ {\it be
an adjoint symmetry of }$X_{F_1\ldots F_{n-1}}${\it . If }$\omega _1,\ldots
,\omega _{n-1}\in \Omega ^1\left( M\right) $ {\it are }$\omega ${\it %
-adjoint pseudosymmetries of }$X_{F_1\ldots F_{n-1}}$ {\it then: } 
\[
{\cal F}=\Lambda \left( \omega ,\omega _1,\ldots ,\omega _{n-1}\right) 
\]
{\it \ is a first integral of }$X_{F_1\ldots F_{n-1}}${\it . In particular,
if }$\omega ,\omega _1,\ldots ,\omega _{n-1}$ {\it are adjoint symmetries
then the above }${\cal F}$ {\it is first integral of }$X_{F_1\ldots F_{n-1}}$%
{\it .}
\end{proposition}

Let us consider a particular case when $\omega ,\omega _i$ are exact $1$%
-forms: $\omega =df,\ \omega _i=df_i$ with $f,f_i\in C^\infty \left(
M\right) $. From the definition of $\Lambda $ we get that ${\cal F}$ given
by Proposition 5.2 is ${\cal F}=\Lambda \left( df,df_1,\ldots
,df_{n-1}\right) =\{f,f_1,\ldots ,f_{n-1}\}$. Recall also that $d{\cal L}_X=%
{\cal L}_Xd$ and then we have:

\begin{proposition} 
{\it Let }$X_{F_1\ldots F_{n-1}}$ {\it be a
Hamiltonian vector field and }\\$f,f_1,\ldots ,f_{n-1}\in C^\infty \left(
M\right) $ {\it such that: }\\(i){\it \ }${\cal L}_{X_{F_1\ldots F_{n-1}}}f$ 
{\it is a closed }$1${\it -form i.e }$d\left( {\cal L}_{X_{F_1\ldots
F_{n-1}}}f\right) ={\cal L}_{X_{F_1}\ldots F_{n-1}}\left( df\right) =0$ \\%
(ii) $L_{X_{F_1\ldots F_{n-1}}}df_i=\rho _idf$ {\it for some} $\rho _i\in
C^\infty \left( M\right) ${\it . Then: } 
\[
{\cal F}=\{f,f_1,\ldots ,f_{n-1}\} 
\]
{\it is a first integral for }$X_{F_1\ldots F_{n-1}}${\it .}
\end{proposition}

It is obvious that this result represents a generalization of Property 3.3
because if $f,f_1,\ldots ,f_{n-1}$ are first integrals then (i) and (ii)
hold with $\rho _i=0$. The next scheme is significant: 
\[
\frame{$
\begin{array}{c}
property \\ 
3.3
\end{array}
$ $
\begin{array}{c}
\{F_1,\ldots ,F_{n-1},f\}=0 \\ 
\{F_1,\ldots ,F_{n-1},f_i\}=0
\end{array}
\rightarrow \{F_1,\ldots ,F_{n-1},\{f,f_1,\ldots ,f_{n-1}\}\}=0$} 
\]
\[
\frame{$
\begin{array}{c}
prop. \\ 
5.3
\end{array}
$ $
\begin{array}{c}
d\{F_1,\ldots ,F_{n-1},f\}=0 \\ 
d\{F_1,\ldots ,F_{n-1},f_i\}=\rho _idf
\end{array}
\rightarrow \{F_1,\ldots ,F_{n-1},\{f,f_1,\ldots ,f_{n-1}\}\}=0.$} 
\]

\begin{trivlist}\item[]
{\bf Example 2.6 revisited.}\ 
For any $f\left( x^1,x^2,x^3\right) $ we have: 
\[
\{F_1,F_2,f\}=x^2x^3\frac{\partial f}{\partial x^1}+x^3x^1\frac{\partial f}{%
\partial x^2}+x^1x^2\frac{\partial f}{\partial x^3} 
\]

We seek a polynomial 
$f=A\left( x^1\right) ^\alpha +B\left( x^2\right)
^\beta +C\left( x^3\right) ^\gamma ,\alpha ,\beta ,\gamma \neq 0$ 
satisfies (i) of Proposition 5.3, that is 
$\{F_1,F_2,f\}=$ constant. Then: 
$A\alpha
\left( x^1\right) ^{\alpha -1}x^2x^3+B\beta x^1\left( x^2\right) ^{\beta
-1}x^3+C\gamma x^1x^2\left( x^3\right) ^{\gamma -1}= x^1x^2x^3\left(
A\alpha \left( x^1\right) ^{\alpha -2}+B\beta \left( x^2\right) ^{\beta
-2}+C\gamma \left( x^3\right) ^{\gamma -2}\right) =$ constant \\ 
which means: 
$A\alpha \left( x^1\right) ^{\alpha -2}+B\beta \left( x^2\right) ^{\beta
-2}+C\gamma \left( x^3\right) ^{\gamma -2}=0$ that is: $\alpha =\beta
=\gamma =2$ and $A+B+C=0$. Therefore: $f=\left( -B-C\right) \left(
x^1\right) ^2+B\left( x^2\right) ^2+C\left( x^3\right) ^2=-2(BF_1+CF_2)$. In
conclusion $f$ is a linear combination of $F_1$ and $F_2$.
\end{trivlist}

\section*{Appendix 1: Remarkable homogeneous NP Ha\-mil\-to\-nian vector
fields}
\stepcounter{section}
\renewcommand{\thesection}{A1}

I) the Euler equations of the free rigid body in ${I\!\!R}^3$ \cite{n:m}, 
\cite{i:np} : 
\[
\dot{x}^1=\left( \frac 1{I_2}-\frac 1{I_3}\right) x^2x^3 
\]
\[
\dot{x}^2=\left( \frac 1{I_3}-\frac 1{I_1}\right) x^3x^1 
\]
\[
\dot{x}^3=\left( \frac 1{I_1}-\frac 1{I_2}\right) x^1x^2 
\]
with
\[
F_1=\frac 12\left[ \frac 1{I_1}\left( x^1\right) ^2+\frac 1{I_2}\left(
x^2\right) ^2+\frac 1{I_3}\left( x^3\right) ^2\right] ,F_2=\frac 12\left[
\left( x^1\right) ^2+\left( x^2\right) ^2+\left( x^3\right) ^2\right] 
\]
is $2$-homogeneous. Let us remark that $F_1$ is {\it the total angular
momentum} and $F_2$ is {\it the kinetic} {\it energy} of the free rigid body.

II) the differential system of Jacobi elliptic functions \cite[p. 137]{c:b}: 
\[
\dot{x}^1=x^2x^3 
\]
\[
\dot{x}^2=-x^3x^1 
\]
\[
\dot{x}^3=-k^2x^1x^2 
\]
with $0<k^2<1$ and: 
\[
F_1=\frac 12\left[ \left( x^1\right) ^2+\left( x^2\right) ^2\right] ,\quad
F_2=\frac 12\left[ k^2\left( x^1\right) ^2+\left( x^3\right) ^2\right] 
\]
is $2$-homogeneous.

III) A more special NP system is given by the 2D isotropic harmonic
oscillator \cite[p. 2]{c:b}: 
\[
\dot{x}^1=y^1 
\]
\[
\dot{x}^2=y^2 
\]
\[
\dot{y}^1=-x^1 
\]
\[
\dot{y}^2=-x^2. 
\]
The following functions $F_1,\ldots ,F_4$ are a basis for the vector space
of all the quadratic integrals of above system \cite[p. 11]{c:b}: 
\[
F_1=x^1y^2-x^2y^1 
\]
\[
F_2=\frac 12\left( x^1x^2+y^1y^2\right) 
\]
\[
F_3=\frac 12\left( \left( y^1\right) ^2+\left( x^1\right) ^2-\left(
y^2\right) ^2-\left( x^2\right) ^2\right) 
\]
\[
F_4=\frac 14\left( \left( x^1\right) ^2+\left( x^2\right) ^2+\left(
y^1\right) ^2+\left( y^2\right) ^2\right) . 
\]
If we define the NP bracket: 
\[
\{f_1,f_2,f_3,f_4\}=\frac 1{F_4}\frac{\partial \left( f_1,f_2,f_3,f_4\right) 
}{\partial \left( x^1,x^2,y^1,y^2\right) } 
\]
then the differential system is given by the Hamiltonian vector field $%
X_{F_1F_2F_3}$ which is $1$-homogeneous.

\section*{Appendix 2: The completeness of Hamiltonian vector fields in
Nambu-Poisson dy\-na\-mics}
\stepcounter{section}
\renewcommand{\thesection}{A2}

We recall that $X\in {\cal X}\left( M\right) $ is {\it complete }if for
every $x_0\in M$ the maximal interval of existence $\left(
t_{-},t_{+}\right) $ for the solution of equation $(2.1)$ with initial
condition $x\left( 0\right) =x_0$ is given by $t_{\pm }=\pm \infty $. A
sufficient condition which assures this property is provided by:

\begin{theorem}[\cite{g:c}] 
{\it Let }$X\in {\cal X}\left(
M\right) ${\it . If there exists }$E,f\in C^\infty \left( M\right) $ {\it %
with }$f$ {\it proper, that is }$f^{-1}\left( compact\right) =compact$, {\it %
\ and }$\alpha ,\beta \in {I\!\!R}$ {\it such that for each }$x\in M$ {\it %
we have: } 
\[
|X\left( E\right) \left( x\right) |\leq \alpha |E\left( x\right) | 
\]
\[
|f\left( x\right) |\leq \beta |E\left( x\right) | 
\]
{\it then }$X$ {\it is complete.}
\end{theorem}

Then we can prove:

\begin{proposition} 
{\it Let }$X_{F_1\ldots F_{n-1}}$ {\it be a NP
Hamiltonian vector field. If there exists }$i\in \{1,\ldots ,n-1\}$ {\it %
such that }$F_i$ {\it is proper} {\it then }$X_{F_1\ldots F_{n-1}}$ {\it is
complete.}
\end{proposition}
\begin{proof}
Let us take in previous theorem $E=f=F_i$. Since $X_{F_1\ldots
F_{n-1}}\left( F_i\right) =0$ it follows that all conditions of the theorem
A2.1 are satisfied. 
\end{proof}

\begin{remark} 
(i){\bf \ }The Poisson case of Proposition A2.2 is Theorem
3.1 from \cite[p. 96]{p:h}. In the  paper cited it is added the assumption
that $F_1$ is bounded below, say $F_1\geq 0$. As Janusz Grabowski point out
in the MR review of \cite{p:h} this assumption is not necessary. \\ (ii)
Also, using a remark of J. Grabowski  in the review cited, we note that it
is sufficient the bracket be just skew-symmetric and satisfying the Leibniz
rule, i.e. be determined by a $n$-vector field.
\end{remark}

\subsection*{Acknowledgement}

I want to thank Prof.\ I.\ Vaisman for providing me with
his preprints on NP brackets. Also, I am very indebted to the referee for
many useful comments.

\label{lastpage}

\end{document}